\documentclass{iau} 
\usepackage{graphicx}

\DeclareUnicodeCharacter{00A0}{ }
\DeclareUnicodeCharacter{00A0}{~}
\usepackage[utf8x]{inputenc}

\title[JD 11. Pre-solar grains and AGB stars] 
{Tracing the role of AGB stars in the Galactic Fluorine budget}

\author[Maryam Saberi ] 
{Maryam Saberi $^{1,2}$
 }

\affiliation{$^1$Rosseland Centre for Solar Physics, University of Oslo, P.O. Box 1029 Blindern, NO-0315 Oslo, Norway \\
$^2$Institute of Theoretical Astrophysics, University of Oslo, P.O. Box 1029 Blindern, NO-0315 Oslo, Norway
 \\ email: {\tt maryam.saberi@astro.uio.no} \\[\affilskip]
}

\pubyear{2021}
\volume{366}  
\jname{The origin of outflows in evolved stars}
\editors{Leen Decin, \& Albert Zijlstra}
\begin{document}

\maketitle

\begin{abstract}
The cosmic origin of fluorine is still under debate. Asymptotic giant branch (AGB) stars are among the few suggested candidates to efficiently synthesis F in our Galaxy, however their relative contribution is not clear. In this paper, we briefly review the theoretical studies from stellar yield models of the F synthesis and chemical equilibrium models of the F-containing molecules in the outflow around AGB stars. Previous detections of the F-bearing species towards AGB and post-AGB stars are also highlighted. We suggest that high-resolution ALMA observations of the AlF, one of the two main carriers of F in the outflow of AGB stars, can provide a reliable tracer of the F-budget in AGB stars. This will be helpful to quantify the role of AGB stars in the Galactic F budget.

\end{abstract}

\section{Introduction}

Fluorine has only one stable isotope ($^{19}$F) whose origin is still uncertain \cite[(e.g. Ryde 2020, Grisoni et al. 2020)]{Ryde20; Grisoni20}. The reason for this lack of understanding is the small number of measurements that have been carried out to observe the abundance evolution of F-containing molecules due to a lack in instrumental sensitivity
or low spectral resolution of the previous generation of observational facilities. F can be easily destroyed by proton, neutron, and alpha particle capture reactions in stellar interiors \cite[(e.g. Ziurys et al. 1994,  Abia et al. 2015)]{Ziurys94; Abia 15}. A number of scenarios have been suggested to explain the origin of Galactic F in literature. The main ones are (i) during thermal pulses and third dredge-up in asymptotic giant branch (AGB) stars with initial masses in a range of $2-4 M_{\odot}$; (ii) during neutrino process occurring in supernova explosions; (iii) during mergers between helium and carbon-oxygen white dwarfs; (iv) during He-burning phase in Wolf-Rayet (WR) stars; (v) in rapidly rotating massive stars \cite[(e.g. Woosley et al. 1995, Meynet et al. 2000, Karakas et al. 2010, Longland et al. 2011, Abia et al. 2015, Jonsson et al. 2017, Limongi et al. 2018, Ryde et al. 2020)]{Woosley95; Meynet00; Karakas10; Longland11; Abia15; Jonsson17; Limongi18; Ryde20}. \cite[Grisoni et al. (2020)]{Grisoni20} has recently published an overview on the F production sites and its impact on the Galactic chemical evolution.
The relative contributions of the suggested sites are still under debate \cite[(e.g. Timmes et al. 1995, Spitoni et al. 2018, Olive et al. 2019)]{ Timmes95; Spitoni18, Olive19}. Among these suggested sites, AGB stars are the only observationally confirmed astrophysical site to efficiently produce F \cite[(e.g. Jorissen et al. 1992, Federman et al. 2005, Werner et al. 2005, Abia et al. 2015)]{Jorissen92; Federman05; Werner05; Abia15}.



\section{Theoretical studies of fluorine in AGB stars}

Theoretical studies suggest that the F production in AGB stars can occur by core and shell He-burning at a temperature of $1.5\times10^8$ K and it can be destroyed once the temperatures exceed $2.5\times10^8$ K.
The synthesized $\rm ^{14}N$ in the hydrogen-burning CNO cycle can produce F by means of a chain of reactions $\rm (^{14}N(\alpha,\gamma) ^{18}F(\beta+) ^{18}O(p, \alpha) ^{15}N(\alpha, \gamma) ^{19}F)$ during the He-burning thermal pulses \cite[(e.g. Forestini et al. 1992; Jorissen et al. 1992; Cristallo et 2014)]{Forestini92; Jorissen92; Cristallo14}. The synthesized F can be then brought to the stellar surface by the 3rd dredge-up and be expelled into the interstellar medium by intense stellar winds or during the planetary nebula phase. 
In high-mass AGB stars with $M > 4M_\odot$, the high temperatures in stellar interiors converts F into Ne. Therefore, AGB stars that can efficiently produce F should be less massive than $4 M_\odot$ to prevent the temperatures of hot bottom burning to destroy the freshly synthesized F \cite[(e.g. Lugaro et al. 2004; Karakas 2010)]{ Lugaro04; Karakas10}.

The results from stellar yield models indicate that F nucleosynthesis is strongly dependent on the stellar mass and metallicity \cite[(e.g. Lugaro et al. 2004; Karakas 2010)]{ Lugaro04; Karakas10}. They predict the highest F production to occur for stars with masses in a range of $M = 2-4 M_\odot$, assuming a metallicity similar to the Solar metallicity.
 Recently updated chemical equilibrium models by \cite[Agundez et al. (2020)]{Agundez20} show that a significant amount of F will be locked into HF and AlF in the inner circumstellar envelope (CSE) of AGB stars within a radius of about $\sim10R_\star$ for all chemical types. Therefore, these molecules are considered to be the best observational tracer of the gas-phase F budget in the outflow of AGB stars.

\section{Detections of F-bearing species in AGB and post-AGB stars}

\cite[Jorissen et al. (1992)]{Jorissen92} presented spectroscopic observations of the infrared vibration-rotation lines of HF towards a sample of AGB stars, showing an over-abundance of $2-30$ times larger than solar F abundance. Estimation of F based on HF lines is subject to large uncertainties due to a large contamination of telluric lines in the same wavelength region. This prevents an accurate determination of the F abundance \cite[(e.g. Abia et al. 2009, 2010 and 2015)]{Abia09; Abia10; Abia15}.
In another study, Werner et al. (2005) reported an over-abundances of $10-250$ times larger than the solar F abundance in a number of hot post-AGB stars from far$-$UV observations of ionized FV and FIV. They suggested the F over-abundance is most likely due to the synthesised F from the preceding AGB phase which is brought to the surface during the post-AGB phase. 

Detection of rotational transitions of several AlF lines are reported towards the well-studied studied carbon-rich AGB star, IRC+10216, by \cite[Ziurys et al. (1994) and Agundez et al. (2012)]{Ziurys94; Agundez12}. 
They reported an average fractional abundance of $f_{\rm AlF/H_2} \sim 10^{-8}$ which is in agreement with the solar abundance of F of $(5\pm2)\times10^{-8}$ that has been recently reported by \cite[Asplund et al. (2021)]{Asplund21}. The initial mass of IRC+10216 is estimated to be $1.6 M_\odot$ by \cite[De Nutte et al. (2017)]{DeNutte17}. Therefore, the fractional AlF abundance is in agreement with predictions from stellar yield models for an AGB star with a solar metallically and initial mass in a range of $1-2 M_\odot$.

In a recently published paper, \cite[Danilovich et al. (2021)]{Danilovich21} reported detections of HF and AlF lines towards the S-type AGB star, W Aql. They have estimated fractional abundances of $f_{\rm HF/H_2}=10^{-7}$ and $f_{\rm AlF/H_2}=4\times10^{-8}$ using radiative transfer analysis. Their reported value of the AlF is higher than expected AlF fractional abundance for W Aql with an initial mass within a range of $1.2-1.6 M_\odot$ reported by \cite[De Nutte et al. (2017)]{DeNutte17}.

In a recent search from ALMA archive data, we have identified detection of AlF line emission towards a sample of five M-type AGB stars (Saberi et al. submitted to A$\&$A). From a rotational diagram analysis of multi-line transitions, we estimated fractional abundances of $f_{\rm AlF/H_2}\sim(2.5\pm1.7)\times10^{-8}$ towards $o$ Ceti and $f_{\rm AlF/H_2}\sim(1.2\pm0.5)\times10^{-8}$ towards R Leo. 
For the rest of sample, we only identified one line which prevents a detailed excitation analysis. We crudely approximation of the AlF fractional abundance to be in a range of $f_{\rm AlF/H_2}\sim(0.1-6)\times10^{-8}$ for W Hya, R Dor, and IK Tau. 
All these sources have an initial mass in a range of $1-2 M_\odot$ estimated by \cite[Decin et al. (2010); Khouri et al. (2014); Hinkle et al. (2016); Danilovich et al. (2017); Velilla Prieto et al. (2017)]{Decin10; Khouri14; Hinkle16; Danilovich17; Velilla17}. Therefore, our results are consistent with the predictions from stellar yield models from \cite[(e.g. Lugaro et al. 2004; Karakas 2010)]{ Lugaro04; Karakas10} and also with the chemical models by \cite[Agundez et al. (2020)]{Agundez20}.

\section{Summary}

We have reviewed the theoretical and observational studies of F, the element whose origin is still under debate, in AGB stars which are the only sources with observational proof to efficiently synthesis F. However, their relative contributions to the total Galactic F budget is still unclear. The results from analysis of new ALMA observations of AlF lines towards a sample of low-mass AGB stars with initial mass in a range of $1-2 M_\odot$ are consistent with theoretical stellar yield models and chemical models (Saberi et al. submitted to A$\&$A).
This study suggest that observations of AlF lines towards AGB stars with initial mass $2-4 M_\odot$ can potentially provide a reliable observational proof of the F nucleosynthesis predicted by stellar yield models and quantity the role of AGB stars in the Galactic F budget. We are granted ALMA observations in Cycle 8 and will perform this analysis in an upcoming paper.

\end{document}